\documentclass[twocolumn,prl,showpacs]{revtex4}
\usepackage{psfrag}
\usepackage{graphicx}
\usepackage{amsmath}
\usepackage{amssymb}

\begin{document}
\title{Dynamical image charge effect in molecular tunnel junctions: Beyond energy level alignment}

\author{Chengjun Jin and Kristian S. Thygesen} \affiliation{Center for
  Atomic-scale Materials Design (CAMD), Department of Physics,
  Technical University of Denmark, DK-2800 Kgs. Lyngby, Denmark.}

\date{\today}

\begin{abstract}
  When an electron tunnels between two metal contacts it temporarily
  induces an image charge (IC) in the electrodes which acts back on
  the tunneling electron. It is usually assumed that the IC forms
  instantaneously such that a static model for the image potential
  applies. Here we investigate how the finite IC formation time
  affects charge transport through a molecule suspended between two
  electrodes. For a single level model, an analytical treatment shows
  that the conductance is suppressed by a factor $Z^2$ (compared to
  the static IC approximation) where $Z$ is the quasiparticle
  renormalization factor. We show that $Z$ can be expressed either in
  terms of the plasma frequency of the electrode or as the overlap
  between the ground states of the electrode with and without an
  electron on the molecule.  First-principles GW calculations for
  benzene-diamine connected to gold electrodes show that the dynamical
  corrections can reduce the conductance by a factor of 2-3.
\end{abstract}

 \maketitle 

The effect of image forces on tunneling electrons was first studied
by Sommerfeld and Bethe\cite{sommerfeld} and Holm\cite{holm} in the
1930s, and later refined by Simmons\cite{simmons} to a form, which
still today is widely used. In Simmons model, the effect of image forces is described
by a $-1/z$ correction to the tunneling barrier. Its range of validity
has recently been critically examined on basis of \emph{ab-initio} calculations and experimental data for (sub-)nanometer sized tunneling junctions\cite{huisman,trouwborst,sanvito,beebe,markussen}. 

Image charge (IC) forces also have important consequences for electron transport at metal-molecule interfaces
because they influence the position of the molecular energy levels relative to the metal Fermi
level\cite{repp05,lu04,kahn03,kubatkin,kaasbjerg,perrin}. Because the
interaction with the image charge lowers the energy cost of adding an
electron/hole to a molecular orbital, the occupied
energy levels are shifted upwards while the empty levels are shifted
downwards in energy as the molecule approaches a metal surface.

Theoretically, the image forces are challenging to describe because
they are created by the electron on which they act. To properly
include such correlation effects one must go beyond standard
single-particle theories like Hartree-Fock and density functional
theory (DFT).\cite{neaton,juanma} For transport in molecular
junctions, this has been done previously using the GW approximation
to the electron
self-energy both in the steady state\cite{gw_transport2,spataru,rangel} and time-dependent\cite{dynamical_image} regimes. Due to
the computational complexity of such many-body methods, simple ad-hoc
correction schemes have been developed which shift the energy of the
molecular orbitals by an amount estimated from a classical image
charge model\cite{bda,mowbray}. Such correction schemes, generally
termed DFT+$\Sigma$, have been shown to improve the agreement with experiments compared
to the uncorrected DFT result\cite{quek}. An interesting question is then whether such a level correction scheme captures all the effects of the IC on electron
transport if the corrections are chosen to reproduce the exact level alignment for the frontier
orbitals. It was recently shown that the IC not only influences the
energy of the molecular orbitals but also their spatial
shape\cite{strange_qp_renorm}. A change in orbital shape will change
the hybridization with the metal states and thereby affect the
tunneling rate. This effect is beyond the DFT+$\Sigma$ schemes, but
should be significant only for highly polarizable molecules.

Except for the few many-body calculations, all previous attempts to
model the IC effect in molecular transport junctions have been based
on the assumption that the IC forms instantaneously. On the other
hand, it is intuitively clear that the role of the IC depends on the
time it takes to polarize the electrode compared to the time the
electron spends on the molecule.  The former is given roughly by the
inverse plasmon frequency of the electrode, $\tau_{p}\approx
1/\omega_p$ while a simple expression for the latter follows from the
time-energy uncertainty relation, $\tau_{\text{tun}}\approx \hbar
/|E_F-\varepsilon_a|$, where $\varepsilon_a$ is the energy of the
molecular orbital closest to the Fermi level.  We note that the
related problem of how the finite plasmon dynamics influences the spatial
form of the image potential at a metal surface has been studied by
several authors in the past\cite{mahan,sunjic,jonson}.

In this paper we show, using both a simple one-level model and
first-principles many-body calculations, that the finite electrode
response time always suppresses the conductance of a molecular
junction compared to the result of non-interacting model with the exact same
level alignment (static IC approximation). Formally this is a consequence
of the reduction of the quasiparticle weight of the molecular
resonance from 1 to $Z<1$ due to the electron-electron interactions
which shift spectral weight from the single-particle excitation to
other excitations (in particular plasmons). In the off-resonant
tunneling regime, the conductance of the one-level model is suppressed
by $Z^2$ compared to the static result. We provide two complementary
physical explanations for this reduction. In a dynamical picture, it
can be related to the ratio between the characteristic IC formation time 
$\tau_p$ and the dwell time of the electron on the molecule 
expressing the reduced screening of the electron due to the ``lagging
behind'' of the IC. In a picture of hopping between many-body states,
$Z$ is given by the overlap of the electrode wave function with and
without the IC and thus explains the origin of the reduced tunneling
rate as a mismatch between the initial and final state of the
electrode.  \emph{Ab-initio} GW calculations for benzene-diamine (BDA)
connected to gold electrodes shows a conductance reduction of almost a
factor three compared to the static approximation (non-interacting
transport through optimally tuned energy levels), demonstrating the
importance of dynamical corrections for realistic systems.

We consider the problem of electron transport through a single electronic level, $|a\rangle$, coupled to left (L) and right (R) electrodes. Due to the
hopping matrix elements between $|a\rangle$ and
the states of the electrodes, the level is broadened into a resonance with a finite spectral width, $\gamma$, which we take to be energy-independent for simplicity. We assume that the level is unoccupied, i.e. $\varepsilon_a>E_F+\gamma$, however, the case of an occupied level is treated completely analogously. The Green's function of the localized level can be written
\begin{equation}\label{eq.gf}
G_a(\omega)= \frac{1}{[\omega-\varepsilon_a-\text{Re}\Sigma_a(\omega)]+i[\gamma+\text{Im}\Sigma_a(\omega)]}
\end{equation}
where the self-energy, $\Sigma_a(\omega) = \langle a|\hat \Sigma(r,r',\omega)|a\rangle$ accounts for the Coulomb interaction between electrons in the electrodes and an electron in $|a\rangle$.
To lowest order in the interaction, the self-energy
contains the Hartree and exchange potentials of
Hartree-Fock theory. These terms do not contribute to the image charge effect and are therefore absorbed in
$\varepsilon_a$. Thus $\Sigma$ includes only the higher order terms (correlation effects).

The screening from the electrodes shifts the pole of the GF from $\varepsilon_a$ to the quasiparticle (QP) energy 
\begin{equation}\label{eq.QP_level}
\varepsilon_a^{QP} = \varepsilon_a + \Delta \varepsilon_{ic}, \quad \Delta \varepsilon_{ic} = Z \cdot \Sigma_a(\varepsilon_a)
\end{equation}
where $\Delta \varepsilon_{ic}$ denotes the image charge shift and $Z=(1-d\Sigma_a(\varepsilon_a)/d\omega)^{-1}$ is the renormalization factor to be discussed later.

\begin{figure}[!t]
\begin{center}
\includegraphics[width=0.8\linewidth]{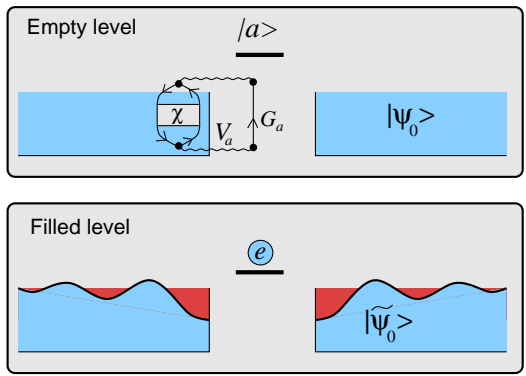}
\caption[system]{\label{fig1} (Color online) Upper panel: When the localized level $|a\rangle$ is empty, the charge distribution corresponding to the electrode ground state, $|\Psi_0\rangle$, is homogeneous (no image charge). Lower panel: When level is occupied, the potential from the localized electron, $V_a(r)$, induces an image charge in the electrode ground state, $|\tilde \Psi_0\rangle$. A Feynman diagram for the self-energy describing the IC effect is shown in the upper panel.}
\end{center}
\end{figure}

Within the GW approximation\cite{hedin}, the self-energy takes the form
\begin{equation}\label{eq.se}
\Sigma(\mathbf r,\mathbf r',\omega) = \frac{i}{2\pi}\int G_0(\mathbf r,\mathbf r',\omega+\omega')\bar W(\mathbf r,\mathbf r',\omega')d\omega'
\end{equation}
where $\bar W = W-v$, and $W$ is the dynamically screened Coulomb
interaction. We have subtracted the bare Coulomb interaction,
$v=1/|\mathbf r- \mathbf r'|$, from $W$ to avoid double counting of the exchange energy which is 
already contained in $\varepsilon_a$. The unperturbed Green's function, is given by
\begin{equation}
G_0(\mathbf r,\mathbf r',\omega) = \frac{\psi_a(\mathbf r)\psi_a(\mathbf r')^*}{\omega-\varepsilon_a+i\gamma}+\sum_{k}\frac{\psi_k(\mathbf r)\psi_k(\mathbf r')^*}{\omega-\varepsilon_k+i0^+\text{sgn}(\varepsilon_k-E_F)}
\end{equation}
In terms of the density response
function of the metal electrode, $\chi$, we have (suppressing the integration over spatial variables) $\bar
W(\omega) = v \chi (\omega) v$. Neglecting the spatial overlap between
$|a\rangle$ and the metal states, the relevant matrix element
of the screened interaction, $\langle a|\bar W(\omega)|a\rangle$, becomes
\begin{equation}\label{eq.wbar}
\bar W_a(\omega) = \int \int V_a(\mathbf r)\chi(\mathbf r, \mathbf r',\omega)V_a(\mathbf r')d \mathbf r d \mathbf r'
\end{equation} 
where $V_a(r)$ is the potential created by an electron in the state $|a\rangle$
\begin{equation}\label{eq.phi}
V_a(\mathbf r) = \int \frac{|\psi_a(\mathbf r')|^2}{|\mathbf r- \mathbf r'|}d \mathbf r'
\end{equation}
A Feynman diagram of the self-energy is shown in Fig. \ref{fig1}.

Using a plasmon pole approximation (PPA) for the response function,  
\begin{equation}
\bar W_a(\omega)= A\Big(\frac{1}{\omega-\omega_p+i\gamma_p}-\frac{1}{\omega+\omega_p-i\gamma_p}\Big )
\end{equation}
the self-energy can be evaluated using complex contour integration
\begin{equation}\label{eq.se2}
\Sigma_a(\omega) = \frac{A}{\omega-\varepsilon_a-\omega_p+i(\gamma+\gamma_p)}
\end{equation}
where $\omega_p$ and $\gamma_p$ are the characteristic plasmon energy and spectral width, respectively. It follows that the imaginary part of $\Sigma_a$ is a Lorentzian of width $\Gamma=\gamma+\gamma_p$ centered at $\omega_p+\varepsilon_a$. In the rest of the paper we assume, for simplicity, that $\Gamma\ll \omega_p$. Since we are only interested in $\Sigma_a(\omega)$ in the range between $E_F$ and $\varepsilon_a$, this means we can set $\Gamma=0$ in Eq. (\ref{eq.se2}). The constant $A$ can be fixed by invoking the condition $\Delta \varepsilon_{ic} = Z \cdot \Sigma_a(\varepsilon_a)$, which results in 
\begin{equation}\label{eq.A}
A= \frac{\Delta \varepsilon_{ic} \omega_p^2}{\omega_p-\Delta \varepsilon_{ic}}.
\end{equation}

Close to equilibrium, i.e. for small bias voltages, the conductance is given by Landauer's fromula $G=\frac{2e^2}{h}T(E_F)$\cite{landauer1}. For the single level model, the transmission at the Fermi level can be written   
\begin{equation}\label{eq.T}
T(E_F) = \frac{\gamma^2}{(E_F-\varepsilon_a^{\text{eff}})^2+\gamma^2}
\end{equation}
where we have defined the effective energy level seen by the tunneling electron as
\begin{equation}
\varepsilon_a^{\text{eff}} = \varepsilon_a + \text{Re}\Sigma_a(E_F) = \varepsilon_a + \Delta \varepsilon_{ic}\Big ( \frac{\omega_p}{|E_F-\varepsilon_a|+\omega_p}\Big ) 
\end{equation}
In the above expression we have assumed, for simplicity of the expression, that $\Delta \varepsilon_{ic} \ll \omega_p$.
The transmission through the interacting level is thus equivalent to transmission through a non-interacting level with energy $\varepsilon_a^{\text{eff}}$. When the image charge formation is fast compared to the average time spent by the electron on the molecule, i.e. when $\omega_p\gg |E_F-\varepsilon_a|$, the effective level equals $\varepsilon_a^{QP}$ and the static image charge approximation is valid. In the opposite regime where the tunneling time is short compared the image charge formation, i.e. $\omega_p\ll |E_F-\varepsilon_a|$, the self-energy vanishes and the tunneling electron ``sees'' the unscreened level $\varepsilon_a$.

In Eq. (\ref{eq.T}) we have embedded the effect of the finite electrode response time into
an effective level position. Although this seems like a reasonable consequence of a partial image charge screening, it does not reflect the correct physics, since
the pole of the Green's function, and thus the spectral peak, remains at
$\varepsilon_a^{QP}$. What is affected is the
renormalization factor, $Z$. Using Eq. (\ref{eq.se2}) and (\ref{eq.A}) it follows that
\begin{equation}\label{eq.Z}
Z = 1-\frac{\Delta \varepsilon_{ic}}{\omega_p}.
\end{equation}
Within the quasiparticle approximation, one expands $\Sigma(\omega)$ to first order around $\varepsilon_a$ which yields the transmission function 
\begin{equation}\label{eq.TQP}
T^{QP}(\omega) = \frac{(Z\gamma)^2}{(\omega-\varepsilon_a^{QP})^2+(Z\gamma)^2}.
\end{equation}
This shows that the transmission resonance remains at
$\varepsilon_a^{QP}$, but is narrowed by a factor of $Z$ compared to
the non-interacting result. In the off-resonant tunneling regime where
$|E_F-\varepsilon_a^{QP}|\gg \gamma$ it follows that the conductance
is suppressed by a factor $Z^2$ compared to the static approximation which sets $\Sigma_a(\omega)=\Delta \varepsilon_{ic}$.

The Green's function formalism describes the propagation of one
electron with the effect of all other electrons of the system is
embedded into the self-energy. Alternatively, one can describe the transport process in terms of transitions between many-body states with different number of electrons on the level.  For non-interacting electrons
this involves only the hopping matrix elements between the state
$|a\rangle$ and the single-particle states of the electrodes,
$|k\rangle$. However, within such a picture we neglect the fact that
all the other electrons in the electrode also feel a change in
potential when the occupation of the localized level changes. To
account for this effect, the single-particle transition matrix element
must be multiplied by the overlap between the initial and final
may-body states of the electrode, $\langle \Psi_0|\tilde
\Psi_0\rangle$. The situation is sketched in Fig. \ref{fig1}.

Using first order perturbation theory to treat the effect of an electron on the molecule, the change in the electrode ground state becomes
\begin{equation}\label{eq.1stpert}
|\Psi^{(1)}_0\rangle  = \sum_{s\neq 0} \frac{\langle \Psi_s|\hat V|\Psi_0\rangle}{E_s-E_0}|\Psi_s\rangle
\end{equation}
where $\hat V = \int \hat n(r) V_a(r)d r$ with  $V_a(r)$ defined in Eq. (\ref{eq.phi}), is the operator describing the potential created by the electron on the level. 

Using the Lehmann representation for the response function in Eq. (\ref{eq.wbar}), performing the integration in Eq. (\ref{eq.se}), and taking the derivative at $\omega=\varepsilon_a$, one obtains
\begin{equation}
Z = \Big (1+\sum_{s\neq 0} \frac{|\langle \Psi_s|\hat V|\Psi_0\rangle|^2}{(E_s-E_0)^2}\Big )^{-1}
\end{equation} 
Noting that the \emph{normalized} final state is $|\tilde \Psi_0\rangle = (|\Psi_0\rangle +
|\Psi^{(1)}_0\rangle)/(1+\langle
  \Psi^{(1)}_0|\Psi^{(1)}_0\rangle)^{1/2}$, and comparing with Eq. (\ref{eq.1stpert}), it follows that
\begin{equation}\label{eq.Z2}
Z=|\langle \tilde \Psi_0|\Psi_0\rangle|^2
\end{equation}
In fact, this also follows from a more general result stating that $Z$ is the norm of the QP state $|a\rangle$ (see e.g. Ref. \cite{strange_qp_renorm}). Eq. (\ref{eq.Z2}) shows that the origin of the $Z^2$ conductance suppression expressed by Eq. (\ref{eq.TQP}) (at least in the co-tunneling regime
where $|E_F-\varepsilon_a^{QP}|\gg \gamma$), can be understood as a mismatch of the initial and final states of the electrodes. Here we note the similarity with
the phenomenon known as Franck-Condon blockade where transport through a molecule is suppressed/blocked due to reduced
overlap between the initial and final vibronic states of the molecule\cite{oppen}. According to Eq. (\ref{eq.Z2}) the magnitude of $Z$ is determined by the relative weight of the component
$|\psi_0^{(1)}\rangle$ in the final state $|\tilde \psi_0\rangle$. We can relate
the norm of $|\psi_0^{(1)}\rangle$ to the response time of the electrode
by noting that the terms in Eq. (\ref{eq.1stpert}) have an $E_s-E_0$
denominator. Within the PPA the dominant terms come from the
plasmon excitations for which $E_s-E_0 \approx \omega_p$. Thus a faster
electrode response, i.e. larger $\omega_p$, is equivalent to a smaller perturbation of the
groundstate and thus $Z$ closer to unity. This is again consistent with Eq. (\ref{eq.Z}).

\begin{figure}[!t]
\begin{center}
\includegraphics[width=1.0\linewidth]{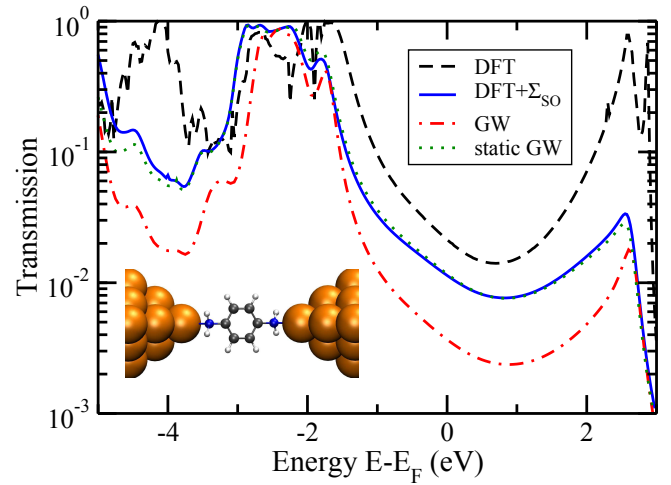}
\caption[system]{\label{fig2} (Color online) The transmission function
  of the gold/BDA junction calculated using four different methods (see text). For the static GW calculations we employed the xc-potential of Eq. (\ref{eq.QPscGW})}
\end{center}
\end{figure}

To test the role of dynamical screening under more realistic conditions, we have performed first-principles GW calculations for the benchmark system of benzene-diamine (BDA) connected to gold electrodes, see inset of Fig~\ref{fig2}.  The details of the calculation follows Ref. \cite{gw_transport2}. In brief, the Green's function of the contacted molecule is obtained by
solving te Dyson equation self-consistently including both lead
coupling self-energies and the GW self-energy. We use a basis of
numerical atomic orbitals at the double-zeta plus polarization level
for the gold electrodes and double-zeta for the BDA. The GW
self-energy is evaluated in a spatial region containing the molecule
and the four closest Au atoms on each side of the molecule. For the considered junction geometry this is sufficient because the IC is essentially confined to the tip Au atoms\cite{gw_transport2}. 

In Fig. \ref{fig2} we show the transmission function calculated using
four different methods. In addition to the GW result we show the
transmission obtained from DFT with the standard PBE xc-functional.
Not surprisingly the latter yields a higher conductance due to the
well known underestimation of the molecular energy gap. To isolate the
role of dynamical effects we have used a "scissors operator" to adjust
the energies of the molecular orbitals in the DFT calculation to those
obtained with GW:
\begin{equation}
\Sigma_{\text{SO}}=\sum_{\nu \in \text{mol}} \Delta \varepsilon_\nu |\psi_\nu\rangle \langle \psi_\nu|
\end{equation}
The molecular orbitals $|\psi_\nu \rangle$ are obtained by diagonalizing the DFT Hamiltonian within the subspace spanned by the basis functions of the BDA. In practice, the energy shift ($\Delta \varepsilon_\nu$) of the three
highest occupied and three lowest unoccupied orbitals are fitted to match the positions of the main peaks in the GW transmission spectrum. As a fourth method we followed the QPscGW scheme of Schilfgaarde \emph{et al.} to construct a static and hermetian xc-potential from the GW self-energy using the expression\cite{qpscgw} 
\begin{equation}\label{eq.QPscGW} 
V^{xc} = \frac{1}{2}\sum_{\nu \mu \in \text{mol}} |\psi_\nu\rangle \text{Re}\{[\Sigma(\varepsilon^{QP}_\nu)]_{\nu \mu } + [\Sigma(\varepsilon^{QP}_\mu)]_{\nu \mu }\} \langle \psi_\mu|
\end{equation}
with the QP energies $\varepsilon^{QP}_\nu$ obtained from the full GW
calculation. As can be seen from Fig. \ref{fig2}, the QPscGW and DFT+$\Sigma_{\text{SO}}$ methods yield very similar transmission spectra. This is because the off-diagonal matrix elements of $V^{xc}$ from Eq. (\ref{eq.QPscGW}) are essentially zero, meaning that the DFT and QP molecular orbitals coincide. (This is not surprising given the low
polarizability of BDA\cite{strange_qp_renorm}). We thus conclude that the observed difference in transmission
between full GW on the one hand and DFT+$\Sigma_{\text{SO}}$ or QPscGW on the other hand, is neither due to differences in energy level alignment nor in the spatial shape of orbitals, but originates from the frequency dependence of the GW self-energy.

\begin{figure}[!t]
\begin{center}
\includegraphics[width=1.0\linewidth]{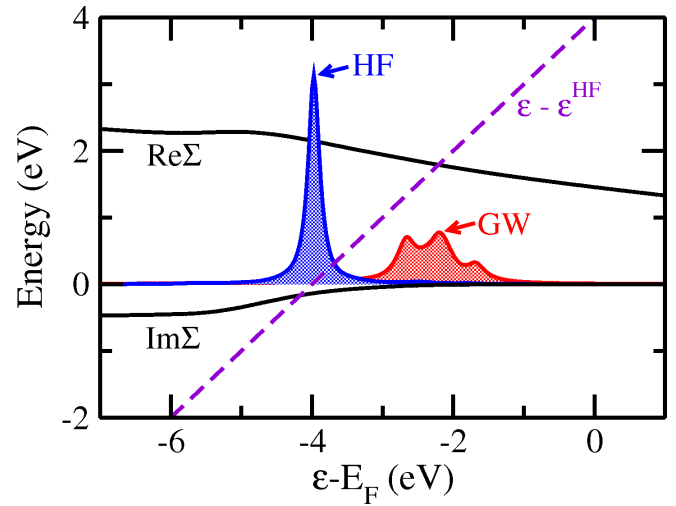}
\caption[system]{\label{fig3} (Color online) The spectral function of
  the HOMO of the contacted BDA molecule calculated with Hartree-Fock
  (blue) and GW (red). The real and imaginary parts of the GW
  self-energy are also shown (black curves).}
\end{center}
\end{figure}

In Fig. \ref{fig3} we show the Hartree-Fock (HF) and GW results for
the spectral function of the BDA HOMO together with the imaginary and real
parts of the GW self-energy $\langle \psi_H|\hat
\Sigma(\omega)|\psi_H\rangle$. From this it follows that the GW
self-energy shifts the HOMO up in energy by 1.9 eV. The corresponding
self-energy shift for BDA in the gas-phase, caused by intra-molecular
screening, is found to be 1.0 eV. From this we conclude that the size
of the IC shift, caused by the metallic screening, is 0.9 eV. It is
clear from the almost linear behavior of $\text{Re}\Sigma(\omega)$,
that the linear expansion of $\Sigma$ leading to Eq. (\ref{eq.TQP}) is well justified.
Furthermore, the imaginary part of the GW self-energy vanishes for
energies above $\varepsilon_H$ in agreement with the one-level model.
The width of the spectral functions in Fig. \ref{fig3} is given by the
imaginary part of the coupling self-energy (not shown). The energy
variation of this broadening follows the density of states at the gold
tip atom. This explains the larger broadening of the GW resonance
compared to the HF resonance which is situated below the gold
$d$-band.

From the slope of $\text{Re}\Sigma$ we obtain the renormalization
factor of $Z=0.84$. Based on the one-level model this should lead to a 
conductance suppression by a factor $Z^2=0.71$ which is, however, not
sufficient to explain the observed difference between the GW and
static GW result, see Fig. \ref{fig2}. The reason for this is that
the BDA junction is not well described by a one-level model. While the
unoccupied states play a minor role for the conductance, the HOMO-2, which
is an anti-bonding version of the HOMO, must be included to obtain a
realistic model. This points to a non-trivial interplay between the dynamical effects on different transport channels.

In conclusion, our results demonstrate that the role of electron-electron interactions in charge transport across a metal-molecule interface goes 
beyond the well established effect on the energy level alignment. 
In general, the image charge dynamics renormalizes the level broadening (or equivalently the tunneling rate) by an amount that depends on the plasmon frequency of the electrode. Since the former can
be tuned, e.g. by nanostructuring or electrostatic gating,
this could provide a basis for experimental investigations of the dynamical image charge effect.

The authors acknowledge support from the Danish Council for Independent Research's Sapere Aude Program through grant no. 11-1051390.


\end{document}